\documentclass[a4paper]{jpconf}
\usepackage{graphicx}
\usepackage{subfigure}
\usepackage{hyperref}
\usepackage{url}

\def\fermi{\emph{Fermi}}

\begin{document}
\title{Extending the \fermi-LAT Data Processing Pipeline to the Grid}
\author{S Zimmer$^{1,2,6}$, L Arrabito$^4$, T Glanzman$^{3,6}$, T Johnson$^{3,6}$, C Lavalley$^4$ and A Tsaregorodtsev$^5$}
\address{$^1$ Department of Physics, Stockholm University, AlbaNova, SE-106 91 Stockholm, Sweden}
\address{$^2$ The Oskar Klein Centre for Cosmoparticle Physics, AlbaNova, SE-10691 Stockholm, Sweden}
\address{$^3$ SLAC National Accelerator Laboratory, Stanford University, Stanford, CA 94305, USA}
\address{$^4$ Laboratoire Univers et Particules de Montpellier, Universit´e Montpellier 2, CNRS/IN2P3, Montpellier, France}
\address{$^5$ Centre de Physique des Particules de Marseille, 163 Av de Luminy Case 902 13288 Marseille, France}
\address{$^6$ on behalf the \fermi-LAT collaboration}
\ead{zimmer@fysik.su.se}

\begin{abstract}
The Data Handling Pipeline ("Pipeline") has been developed for the \fermi~Gamma-Ray Space Telescope (\fermi)~Large Area Telescope (LAT) which launched in June 2008. Since then it has been in use to completely automate the production of data quality monitoring quantities, reconstruction and routine analysis of all data received from the satellite and to deliver science products to the collaboration and the Fermi Science Support Center. Aside from the reconstruction of raw data from the satellite (\emph{Level 1}), data reprocessing and various event-level analyses are also reasonably heavy loads on the pipeline and computing resources.  These other loads, unlike Level 1, can run continuously for weeks or months at a time. In addition it receives heavy use in performing production Monte Carlo tasks. 

In daily use it receives a new data download every 3 hours and launches about 2000 jobs to process each download, typically completing the processing of the data before the next download arrives. The need for manual intervention has been reduced to less than 0.01\% of submitted jobs.

The Pipeline software is written almost entirely in Java and comprises several modules. The software comprises web-services that allow online monitoring and provides charts summarizing work flow aspects and performance information. The server supports communication with several batch systems such as LSF and BQS and recently also Sun Grid Engine and Condor. This is accomplished through dedicated job control services that for \fermi~are running at SLAC and the other computing site involved in this large scale framework, the Lyon computing center of IN2P3. While being different in the logic of a task, we evaluate a separate interface to the Dirac system in order to communicate with EGI sites to utilize Grid resources, using dedicated Grid optimized systems rather than developing our own. 
More recently the Pipeline and its associated data catalog have been generalized for use by other experiments, and are currently being used by the Enriched Xenon Observatory (EXO), Cryogenic Dark Matter Search (CDMS) experiments as well as for Monte Carlo simulations for the future Cherenkov Telescope Array (CTA).
\end{abstract}

\section{The \fermi-LAT mission}
The Fermi spacecraft supports two gamma-ray instruments; the Large Area Telescope (LAT) \cite{atwood} and the Gamma-ray Burst Monitor (GBM) \cite{gbm}. The LAT is a wide-field gamma-ray telescope (20 MeV - 300 GeV) that continuously scans the sky, providing all-sky coverage every two orbits. The GBM is an all-sky monitor (10 keV - 25 MeV) that detects transient events such as occultations and gamma-ray bursts (GRB). GBM detections of strong GRBs can result in an autonomous re-point of the observatory to allow the LAT to obtain afterglow observations.
The satellite sends data to the ground every 3 hours. Data is transferred via relay satellites at 40 MB/s to the White Sands ground station. It then follows a leased line to the Mission Operations Center (MOC) at Goddard Space Flight Center where data is split into two parts and sent for science processing to both the GBM and the LAT teams, the latter being located at Stanford National Accelerator Laboaratory (SLAC). 
\section{Computing Requirements and Principle Task Types}
The processing of downlinked satellite data is a time-critical operation. It is, therefore, necessary to automatically trigger the Pipeline for each newly arrived block of data, and to exploit parallel processing in a batch farm to achieve the required latency for the production of the various data products.  This processing is complex and is abstracted in a process graph.  An xml representation of this process graph is interpreted by the Pipeline to become a \emph{task}.  Once defined, a task is exercised by the creation of \emph{streams}, each of which is one instance of the process graph and which consists of an arbitrary number of interconnected batch jobs (or scripts) known as \emph{process instances} \cite{flath}. To that end, the software is designed to be able to handle and monitor thousands of streams being processed at separate sites with a daily average throughput of about 1/2 CPU-year of processing. To date peak usage has been 45,000 streams in a single day and 167 CPU-years of processing in a single month \cite{dubois}. 

The Pipeline was designed with the L1 data processing task (``level 1 processing'', the core data processing of raw data that comes from the satellite), automatic science processing (ASP) and Monte Carlo simulations as principle task types in mind and to ensure the tight connection to the Fermi Data Catalog. It is literally impossible to depict the whole L1 task scheme in one single figure as it contains many dozens of sub-stream creations, dependencies and automatic re-run mechanisms. Thus we refrain from including them in this paper. For details the reader is referred to \cite{focke}.
\begin{figure}
\begin{center}
\includegraphics[width=14cm]{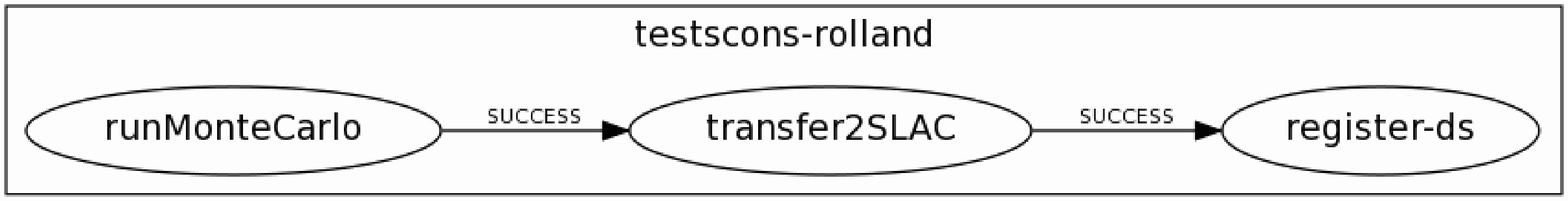}
\caption{Graph for a typical MC task. Each task contains three steps: \emph{runMonteCarlo}, \emph{transfer2SLAC} and executed from the Pipeline server \emph{register-ds} which executes a Jython scriptlet registering the data proucts received in the Fermi Data Catalog. Since failures of streams for this task often require user intervention, we refrain from implementing an automatic roll-back if the first step fails.}
\label{fig:MCTask}
\end{center}
\end{figure}

Instead we show the simple layout of a Monte Carlo task (as our efforts initially are geared towards porting them to Grid sites) in Fig.~[\ref{fig:MCTask}] that does not rely on external dependencies such as local databases. This task simply consists of 3 steps, the generation of Monte Carlo data on a computing node, the transfer of the MC products to SLAC and their registration in the Data Catalog. The first 2 steps are batch operations where the registration step takes only split seconds and is achieved through running a dedicated Jython \cite{jython} scriptlet. 

\section{Pipeline Components}
\begin{figure}[htp]
\begin{center}
\subfigure[Pipeline Architecture and Key Technologies]{\label{fig:keytech}\includegraphics[width=.8\textwidth]{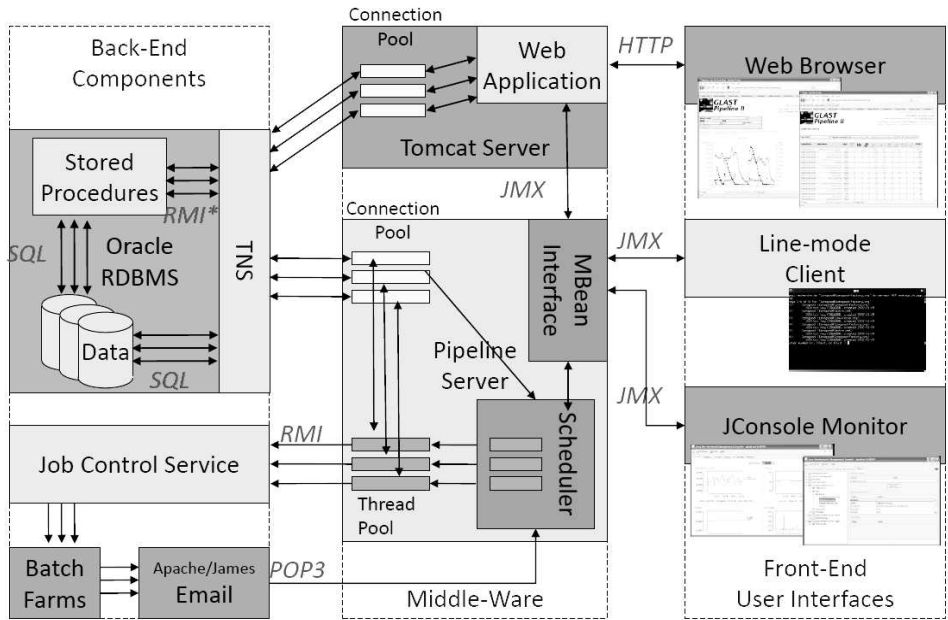}}
\subfigure[Implementation of Job Control Services and Data Catalog]{\label{fig:jcd}\includegraphics[width=.8\textwidth]{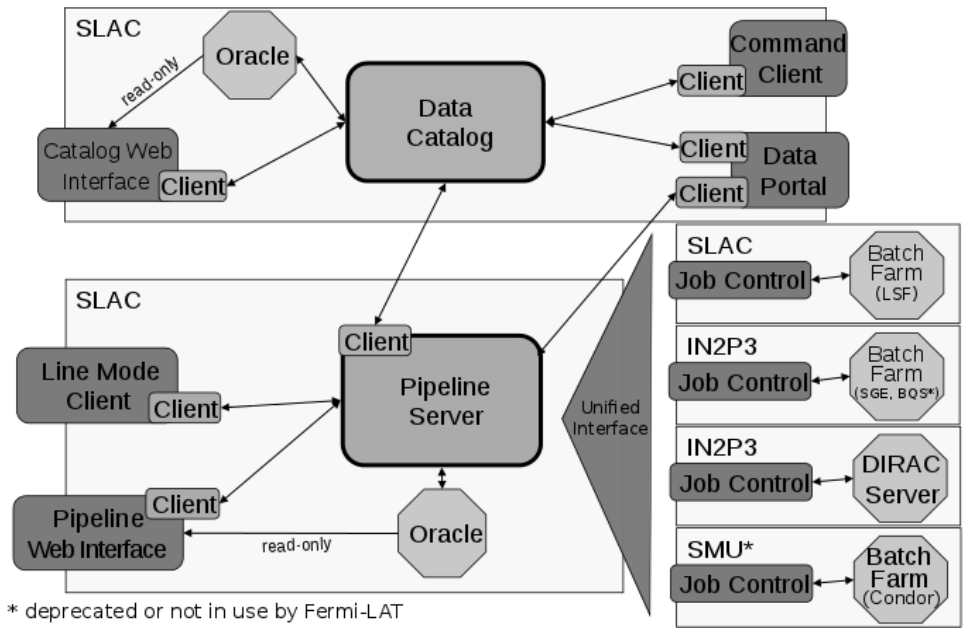}}
\end{center}
\caption{Pipeline architecture and key technologies (a) and Connection of Pipeline Server with existing Job Control Services (b). To date the Pipeline supports interfaces to BQS, LSF, Condor, SGE and the DIRAC server that has been configured for the use with \fermi.}
\label{fig:p2components}
\end{figure}

We use a three tier architecture as shown in Fig.~[\ref{fig:keytech}] comprising of back-end components, middle-ware and front-end user interfaces. We describe them more in detail below.
\subsection{Back-End components}
Core of the back-end components is the Oracle Database that stores all processing states. In addition we make extensive use of Oracle technologies such as the scheduler that is used to run periodic jobs for system monitoring including resource monitoring of the Oracle server itself. Most quantities are made available to the user through trending plots that allow quick judgment about the state of the system along with its resource usage. Another back-end component is the Pipeline Job Control Service. Using Remote Method Invocation (RMI \cite{rmi}) they publish the uniform interface and communicate with the Pipeline server. We discuss some more details on this component in the next section. 

To provide asynchronous persistent messaging from batch jobs to the Pipeline server we use email messaging. At the beginning of a job an email is sent detailing the host name and other worker node specific information. Another email is sent to indicate that a job has finished. This email may also contain additional commands that invoke new Pipeline commands, such as creation of sub-streams as the next step. In order to avoid overloading the SLAC email server, by relaying tens of thousands emails per day, we use a dedicated email server running the free Apache JAMES \cite{james} software. 
\subsection{Middleware}
The Pipeline server is the core of the system and contains two pools for threads, a worker and an admin pool. When a batch process is ready to run on one of the farms, a thread is allocated on the worker pool to perform the submission using the appropriate Job Control Service. Extensions to the pipeline (called \emph{plugins}) can be used to add additional functionality, for example to provide access to experiment specific databases or to communicate with other middleware services without compromising the experiment independent design of the core Pipeline software. Plugins are written in Java and loaded dynamically when the Pipeline starts. The user can also provide Jython scriptlets to run within the Pipeline threads to perform simple calculations and to communicate with plugins. The Fermi data catalog is implemented as a plugin. 

The Pipeline server API allows queries for processes, stream management as well as means to get or set environment variables. The admin thread pool is used to identify work to be delegated to the worker pool. This includes gathering processes which are ready to run as well as various database queries. 

\subsection{Front-End Components}
We provide a subset of the Pipeline API as Java Management Extension (JMX), that provides a call interface to various user-interface applications. These come both as Web interface and line command applications. The web interface provides password protected world-wide access to the Pipeline and its control interfaces and allow simultaneous monitoring of tens of thousands of jobs in various states. For detailed technical information on the Pipeline components the reader is referred to \cite{flath}.

\subsection{The Pipeline Job Control Service} 
Each Job Control Service (refer to Fig.~[\ref{fig:jcd}]) implements job control and status methods that are specific to the batch or Grid system. To that end each Job Control Service needs to provide the following commands: \emph{SubmitJob, GetJobStatus, KillJob}. The code for the Job Control Service is written in Java and runs as a daemon on a dedicated service machine at each computing site. To date the Pipeline supports LSF \cite{hpc}, BQS and more recently also Condor as well as Sun Grid Engine (SGE \cite{sge}). At present \fermi~uses LSF at SLAC and SGE at the Lyon computing center. The code can easily be adapted to any other desired batch system that follows the same job logic as the currently supported ones. To that end a java class \emph{BatchJobControlService} and \emph{BatchStatus} need to be implemented for the desired new batch system, where \emph{Batch} denotes the system to be implemented. 

For the use with DIRAC (Distributed Infrastructure with Remote Agent Control) the Job Control code wraps python scripts that provide the bridge in both format and language by using the dedicated DIRAC API. We describe further aspects of the DIRAC system and motivate our choice to use it in Section \ref{DIRAC}.

In the past a task was defined on top level to be handled by a specific Job Control Service. Recently the Pipeline has been enabled to support multi-site tasks that allow the Pipeline server to delegate the running of commands to our DIRAC Job Control Service while we leave the transfer step to the Lyon-based SGE Job Control Service.
\section{Integrating the Pipeline with the Grid}
As of this writing the LAT has been granted resources both at SLAC (dynamic allocation scheme) and Lyon (guaranteed 1200 cores allocation). At Lyon all resources are used for Monte Carlo productions while at SLAC the total allocation is shared between L1, Monte Carlo production and individual user jobs from the collaboration. As a recent challenge we have started reprocessing all of our data that was taken from the beginning of the mission up until now and are reprocessing it with our current state of knowledge about the experiment. The reprocessing requires a significant amount of our computing resources. As allocations are dynamic, users running science analysis may be directly impacted through less available slots on the batch farm. Another recent challenge was a massive Monte Carlo production of proton runs that occupied our resources both at SLAC and Lyon for several months. While not being particularly storage intensive, we do require significant amounts of CPU time. Since this production run was setup with the previous interation of the instrument response, dubbed ``pass 6'', it is likely that simulation requests of this kind may be repeated as our knowledge of the experiment grows. It is thus important to investigate possibilities to extend our resources that can be utilized within the current Pipeline framework.
\subsection{The glast.org VO}
Although we perform standard computing tasks at our two sites on local batch farms, there exists the virtual organization (VO) glast.org. This organization was founded in 2009 to provide access to gLite \cite{glite} resources granted by participating institutions in Italy and France. At present the VO includes 13 sites that are partially enabled for use by \fermi. Its use has however been limited to non-pipeline operations. Most notably, our existing Grid resources have been used for stand-alone pulsar blind searches and some large Monte Carlo simulations. These stand-alone tasks were unable to take advantage of the pipline's integration with the batch system and data catalog which made them more man-power intensive than their Pipeline counterparts.
\subsection{DIRAC as potential connection to EGI enabled sites\label{DIRAC}}
In order to optimize the resource usage provided by the EGI \cite{egi} sites supporting the glast.org VO, we are exploring the use of the DIRAC system as potential connection between the Pipeline and the Grid resources \cite{tsaregodtsev}.

The DIRAC system, originally developed to support production activities of the LHCb experiment, is today a general solution to manage the distributed computing activities of several communities. One of its main components is the Workload Management System (WMS). The key feature of the DIRAC WMS is the implementation of the ‘pilot job’ mechanism, which is widely exploited by all LHC communities as a way to guarantee high success rate for user jobs (workloads). More in detail, workloads are submitted to the DIRAC WMS and inserted in the DIRAC \emph{Central Task Queue}. The presence of workloads in the DIRAC \emph{Central Task Queue} triggers pilot jobs submission. Pilot jobs are regular Grid jobs that are submitted through the standard gLite WMS. Once the pilot job gets on the worker node, it performs some environment checks and only in case these tests succeed, the workload is pulled from the DIRAC \emph{Central Task Queue} and executed. In case the environment checks fail or the pilot job gets aborted (for example because of a mis-configured site), only the pilot job is affected. The net result of this mechanism is a significant improvement on the workload success rate. Also, since the resources are pre-reserved by pilot jobs, the waiting time to get the workload execution started is reduced.

\subsection{Establishing an Interface between Pipeline and DIRAC: Design considerations}
\begin{figure}
\begin{center}
\includegraphics[width=14cm]{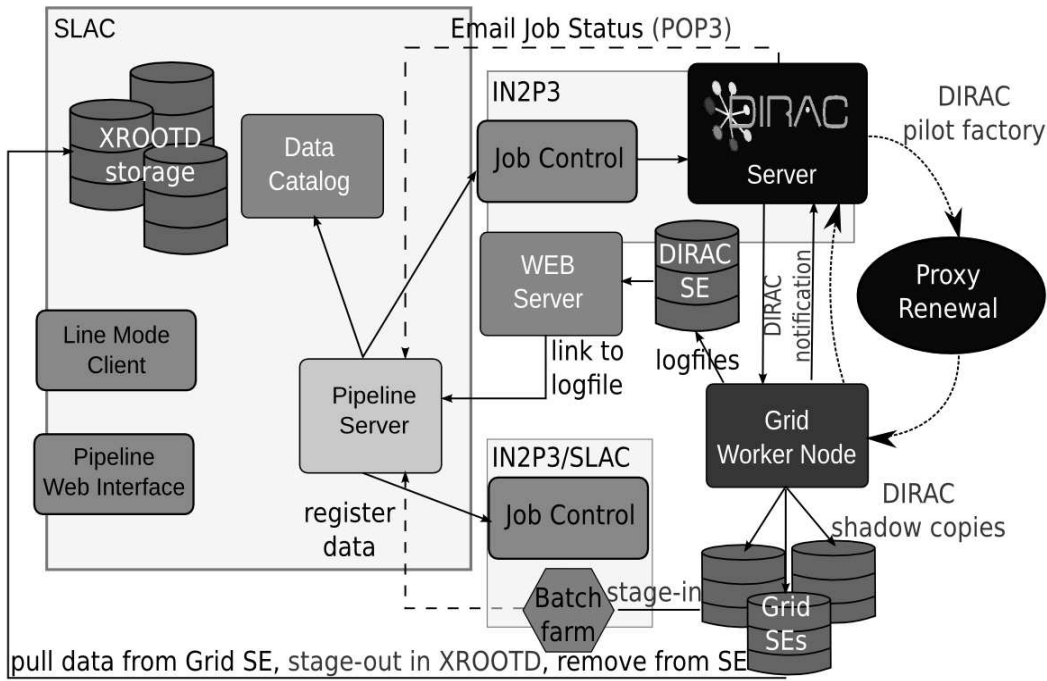}
\caption{Our proposed integration of the DIRAC system with the Pipeline. We establish an additional Job Control Daemon that delegates job requests to the DIRAC server. This server maintains its communication to the worker node through a dedicated notification service. Log files are stored on a DIRAC storage element and accessed via a web server.}
\label{fig:p2dirac}
\end{center}
\end{figure}
While DIRAC has specifically been designed to tackle intrinsic Grid inefficiencies through its pilot job concept, it cannot solve some of the initial issues when being connected to the Pipeline system. In particular:
\begin{itemize}
\item{The Grid uses personalized certificates. When submitting jobs through the Pipeline web interface, we submit them with a generic user ID, which is most feasible as the number of authorized Pipeline users is small.}
\item{Some of our sites, in particular those from INFN, cannot directly send emails to the Pipeline server, thus rendering the standard communication of job status and task logic de-facto unusable.}
\item{In general large data on the Grid would be stored on Grid Storage Elements (Grid SE) and require user intervention to download it once finished. This makes the automatic copy to our central SLAC XROOTD space difficult to be used, in particular using the tight integration with the data catalog.}
\end{itemize}
We decided to implement an interface to the DIRAC system for several reasons:
\begin{itemize}
\item{Independence of Grid middleware: Currently the VO comprises only gLite sites but DIRAC is able to communicate with all common Grid middleware platforms, thus making it less difficult to connect to other Grid initiatives, such as the Open Science Grid in the US \cite{osg}.}
\item{Additional Monitoring functionality: DIRAC introduces a more detailed job monitor that in addition reports minor and major application statuses together with the overall job status, thus allowing easier debugging of tasks without the definite need to inspect log files manually. This task has always been among the most time consuming. We hope to achieve an improvement by using the new monitor, albeit as read-only system. The reason for that is that the Pipeline ID and the Grid ID are generally not the same and a re-submitted job on the Pipeline keeps its ID while on the Grid it is relaunched and assigned a new unique Grid ID.}

\item{While \fermi~is entering its 4th mission year, the number of developers for software has begun to dwindle. Thus it is important to keep the required manpower as low as possible while maintaining full performance and possibly improving the capabilities of the Pipeline. By using DIRAC we can build our interface on a system in active development with many possibilities to influence the development process to meet our needs.}
\end{itemize}
Implementing the DIRAC interface comes in two steps, first the Job Control Daemon as described in the previous section that wraps the basic Pipeline commands through the DIRAC API and secondly the configuration of a dedicated DIRAC server. The implementation is shown in Fig.~[\ref{fig:p2dirac}].

We make use of two of DIRACs core technologies: the pilot factory mechanism is used to renew proxies and authenticate the Job Control Daemon to the Grid allowing us to use one certificate retaining our previous user scheme. The DIRAC notification service provides means for the DIRAC server to communicate with Grid worker nodes. This service is both safe to inception and can be modified to relay the content of our status emails to the DIRAC server that itself implements the email communication with the Pipeline server. 

One typical by-product of running our code are more or less extensive log files. Usually we have several hundreds of small log-files that each do not exceed a few MB. In the past other experiments had to artificially enlarge their log files to ensure the stability of the Grid SEs due to journalling of small files. We can effectively circumvent this by declaring a local storage element at the computing center in Lyon as a dedicated DIRAC SE. These storage elements are addressed like normal Grid elements but they do not need journaling to function. Since this is a local SE, we can view its content e.g. through a Web server providing access to log-files that can directly be linked from the Pipeline server. 

In conclusion we believe that this solution provides an easy to implement and maintain interface to the Pipeline system used for \fermi-LAT. 

\section{Conclusions}
The existing VO resources of glast.org suggest the possibility to establish a new connection to Grid services with our existing Pipeline architecture. We mitigate issues such as the transfer and subsequent registration of data products at the Fermi Data Catalog by using existing Pipeline technologies. Grid inefficiencies are handled by the DIRAC system that acts as a broker providing asynchronous communication with Grid worker nodes and a closed mechanism to automatically renew proxies to use for Grid operations. We leave handling of meta data to the existing Pipeline technologies. 

As such, the Pipeline software itself was designed in a manner to not contain any Fermi specific functionality. Through the plug-in feature of the middleware it is successfully used for data handling for the Enriched Xenon Observatory or MC simulations for the Cryogenic Dark Matter Search experiment (CDMS \cite{cdms}), the upcoming Cherenkov Telescope Array (CTA \cite{cta}) and the Large Synoptic Survey Telescope (LSST \cite{lsst}), and is also being considered for use by other NASA missions.

\ack
We acknowledge the ongoing generous support of SLAC (US) and IN2P3 (France) as well as we thank them for 
increasing allocations at both sites. Furthermore we are thankful for Grid resources provided by INFN and IN2P3.

SZ likes to thank the organizers for a stimulating meeting and particularly helpful discussions with St\'{e}phane Guillaume Poss.

The \fermi-LAT Collaboration acknowledges generous ongoing support
from a number of agencies and institutes that have supported both the
development and the operation of the LAT as well as scientific data analysis.
These include the National Aeronautics and Space Administration and the
Department of Energy in the United States, the Commissariat \`a l'Energie Atomique
and the Centre National de la Recherche Scientifique / Institut National de Physique
Nucl\'eaire et de Physique des Particules in France, the Agenzia Spaziale Italiana
and the Istituto Nazionale di Fisica Nucleare in Italy, the Ministry of Education,
Culture, Sports, Science and Technology (MEXT), High Energy Accelerator Research
Organization (KEK) and Japan Aerospace Exploration Agency (JAXA) in Japan, and
the K.~A.~Wallenberg Foundation, the Swedish Research Council and the
Swedish National Space Board in Sweden.

Additional support for science analysis during the operations phase is gratefully
acknowledged from the Istituto Nazionale di Astrofisica in Italy and the Centre National d'\'Etudes Spatiales in France.

\end{document}